# Fragile electron-phonon superconductivity in $MnB_4$ under pressure

Renhai Wang[1#], Shiya Chen[2#], Feng Zheng[3†], Zhen Zhang[4], Xingzhi Wang[2], Huafeng Dong[1], Cai-Zhuang Wang[4,5], Vladimir Antropov[4,5], Kai-Ming Ho[4], Yang Sun[2*]

[1]*School of Physics and Optoelectronic Engineering, Guangdong University of Technology,* Guangzhou *510006, China*
[2]*Department of Physics, Xiamen University, Xiamen 361005, China*
[3]*School of Science, Jimei University, Xiamen 361021, China*
[4]*Department of Physics and Astronomy, Iowa State University, Ames, Iowa 50011, USA*
[5]*Ames National Laboratory, U.S. Department of Energy, Ames, Iowa 50011, USA*

The origin of pressure-induced superconductivity in $MnB_4$ remains unclear. Here we show that it can be explained by electron-phonon coupling once the structural space is mapped using both volume and the Mn dimer distance as key structural parameters under compression. Minor changes in the dimer distance significantly affect electronic and phonon properties, bringing the calculated $T_c$ into agreement with experiment. Our results suggest that $MnB_4$ is a highly responsive system, providing a platform for probing the subtle interplay between structural instability, superconductivity and magnetism.

Superconductivity is a highly fragile quantum state that relies on the coherent pairing of electrons. This coherence can be easily disrupted by many factors, among which structural distortions and strain are especially common. For instance, the recent discovery of high-temperature superconductivity in bulk $La_3Ni_2O_7$ under pressure suggests that subtle changes in Ni-O bonding play an important role in stabilizing superconductivity [1]. In kagome metals such as the $AV_3Sb_5$ family (A = K, Rb, Cs), superconductivity is also highly sensitive to structure [2]: local strain can significantly enhance critical temperature $T_c$ and modify the gap behavior, while high pressure suppresses the charge-density wave and creates a new superconducting state. These systems can be characterized as highly responsive systems (HRSs), when a physical system is in a state characterized by an order parameter that is very sensitive to the external perturbations. Such HRS behavior is naturally related to proximity to structural or electronic instabilities that can promote superconductivity [3].

Recent experiments reported pressure-induced superconductivity in $MnB_4$, with $T_c$ reaching 14 K near 150 GPa [4]. Density-functional perturbation theory calculations for this system yield $T_c$ < 1 K at high pressure [4] and rule out a conventional electron-phonon coupling (EPC) mechanism of superconductivity. This has led to speculation that the superconductivity may originate from an unconventional pairing mechanism from spin fluctuations [4]. Unlike high-symmetry superconducting borides such as $MgB_2$, in $MnB_4$, a Peierls distortion drives a structural transition from the orthorhombic phase (typical of similar $XB_4$ borides) to a lower-symmetry monoclinic crystal structure. This lattice instability triggers a strong dimerization of one-dimensional Mn-Mn chains, resulting in unusually short Mn-Mn distances. The distortion opens a pseudo-band gap at the Fermi energy, pushing $MnB_4$ toward semiconducting behavior, while its proximity to an itinerant state preserves complex magnetic properties and significant physical anomalies [5-7]. The superconducting mechanism remains a mystery, as the strong suppression of spin fluctuations under pressure is expected.

In this work, we examine the low-symmetry monoclinic structure of $MnB_4$ in greater detail, which points to a missing structural factor in earlier EPC calculations. We will demonstrate below that $MnB_4$ can be considered an HRS with fragile EPC-induced superconductivity that is highly susceptible to the Mn-Mn bonding distance. This sensitivity is unusually dramatic, as changes within the experimental uncertainty can strongly modify $T_c$. Our results show that the $T_c$ observed near ~140 GPa can be fully explained by an EPC mechanism once the experimentally relevant Mn-Mn distances are considered. The Mn-Mn distance can therefore be viewed as a key tuning parameter for superconductivity in $MnB_4$ under pressure.

***Structural landscape.***—We first compare the structural motif of $MnB_4$ with those of the higher-symmetry borides $MgB_2$ and $FeB_4$ in Figs. 1(a)-1(c). The hexagonal P6/mmm structure of $MgB_2$ contains regular boron hexagons and a uniform hexagonal metal layer, with identical Mg-Mg separations of ~3.1 Å (Fig. 1a). $FeB_4$ has an orthorhombic Pnnm structure with half-vacant metal layers. The remaining

[#]Equal contribution.
[†]Email: fzheng@jmu.edu.cn
[*]Email: yangsun@xmu.edu.cn

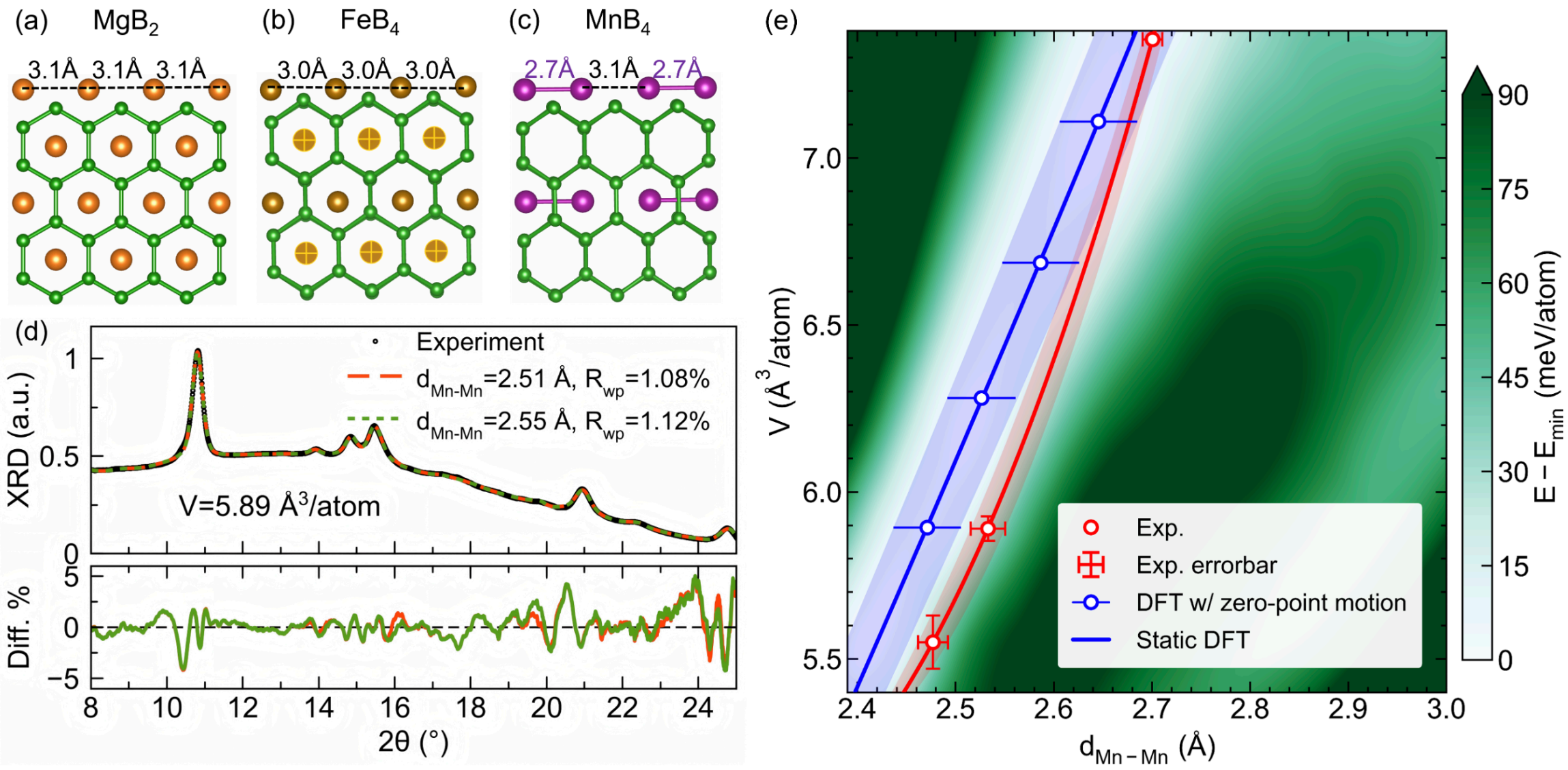

**FIG. 1** Mn-Mn dimerization and two-parameter energy landscape in $MnB_4$. (a) *P6/mmm* $MgB_2$ structure with regular B hexagons and a uniform Mg-Mg chain with a distance of 3.1Å. (b) Orthorhombic *Pnnm* $FeB_4$ structure with Fe atoms form an undimerized one-dimensional metal chain with equal Fe-Fe separations of ~3.0Å. Crossed and uncrossed Fe atoms denote two layers occupying distinct interstitial positions between neighboring boron layers. (c) Monoclinic $P2_1/c$ $MnB_4$ structure with Mn chain dimerized into alternating short and long Mn-Mn bonds, with representative distances of ~2.7 and ~3.2Å, respectively. (d) Representative refinements of the experimental XRD pattern at 110 GPa using the $MnB_4$ structure with the same volume of 5.89 Å$^3$/atom but different Mn-Mn dimer distances, $d_{\text{Mn-Mn}}$=2.51Å and 2.55 Å, yielding weighted residuals ($R_{wp}$) of 1.08% and 1.12%, respectively. (e) The volume $V$ and Mn-Mn dimer distance $d_{\text{Mn-Mn}}$ for monoclinic $MnB_4$. The color map shows the energy difference relative to the lowest-energy structure at fixed V. Blue symbols mark the static 0 K DFT equilibrium structures, with horizontal bars indicating the local harmonic vibrational range arising from zero-point energy. Red symbols denote the XRD-refined experimental structural data, with error bars representing the confidence interval of $d_{\text{Mn-Mn}}$ obtained from the XRD refinement. The lines are quadratic fits.

Fe atoms still form a uniform chain with equal Fe-Fe separations of ~3.0 Å (Fig. 1b). In both cases, the higher lattice symmetry constrains the metal sublattice, so that a finite alternation of metal-metal distances would require additional symmetry lowering. By contrast, $MnB_4$ adopts a monoclinic $P2_1/c$ structure with much lower symmetry (Fig. 1c). Along its quasi-one-dimensional Mn chain, the Mn-Mn separation can modulate without introducing additional symmetry breaking, allowing alternating short and long Mn-Mn bonds with distances of 2.7 and 3.1 Å under ambient conditions, respectively. Thus, the relevant structural response is not described by volume alone. The Mn-Mn dimer distance $d_{\text{Mn-Mn}}$ provides an additional structural degree of freedom under compression.

To determine the experimentally observed $V$-$d_{\text{Mn-Mn}}$ relation, we refined the $MnB_4$ structures at 110 and 158 GPa using previously reported powder XRD patterns and the GSAS package [8] and the EXPGUI interface [9], with a weighted-profile residual criterion of $R_{wp}$ < 2% (see Supplemental Material, Text S1). Within this typical criterion, the refined Mn-Mn bond distance can vary by ~0.1 Å. Figure 1(d) compares the measured XRD pattern at 110 GPa with calculated XRD profiles for two structures with the same volume but slightly different $d_{\text{Mn-Mn}}$ values of 2.51 and 2.55 Å. Both structural models fit the XRD data well, demonstrating that small variations in $d_{\text{Mn-Mn}}$ are allowed by the experimental refinement. Similar results are observed for the 158 GPa data, as shown in Supplemental Material Fig. S1. We therefore plot the experimentally refined structures in the $V$-$d_{\text{Mn-Mn}}$ map in Fig. 1(e), with confidence intervals indicating possible variations in $d_{\text{Mn-Mn}}$.

We compute the $V$-$d_{\text{Mn-Mn}}$ relation by optimizing the $MnB_4$ structure using DFT within the generalized gradient approximation (GGA), which has been shown to give a reliable overall description of the relative energies and electronic structure of $MnB_4$ [7]. Because DFT-calculated pressures can be sensitive to exchange-correlation effects [10, 11], we compare theory and experiment at fixed volume rather than by nominal pressure. The computational details are given in the Methods, and the calculated $V$-$d_{\text{Mn-Mn}}$ curve is shown in Fig. 1(e). By varying $d_{\text{Mn-Mn}}$ at fixed volume, we compute the energy landscape in the $V$-$d_{\text{Mn-Mn}}$ space, shown as the color map in Fig. 1(e). The landscape contains a broad low-energy valley near the optimized structure, where different $d_{\text{Mn-Mn}}$ values have very similar energies, within a few meV. Such a flat energy valley can lead to appreciable variation in $d_{\text{Mn-Mn}}$ associated with the Peierls distortion. A harmonic estimate shows the zero-point displacement of

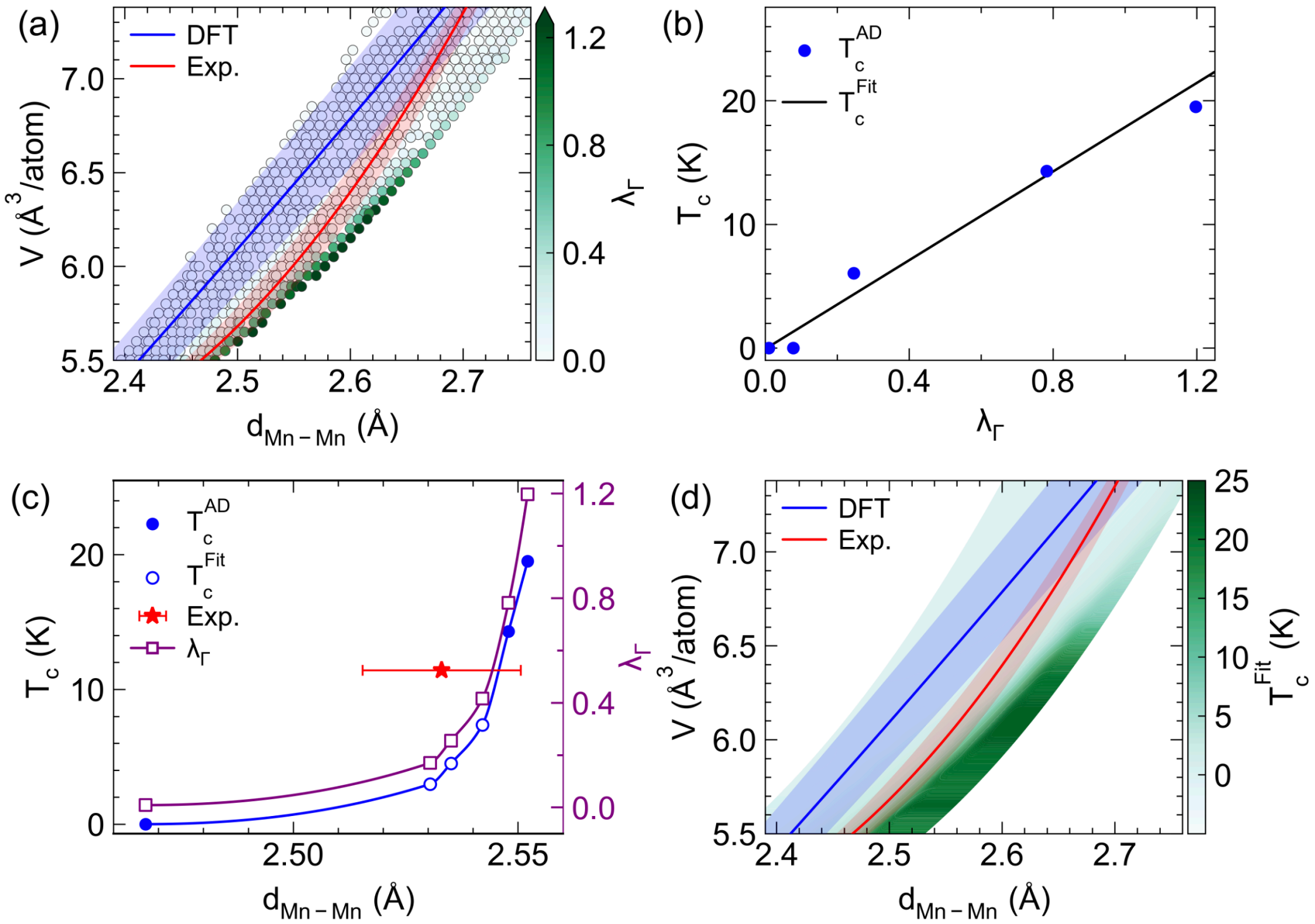

**FIG. 2** Electron-phonon coupling and superconductivity landscape in $MnB_4$. (a) $\lambda_\Gamma$ in the $V$–$d_{\text{Mn-Mn}}$ plane, constructed from 561 sampled structures. The blue curve denotes the DFT relaxed structure with the shaded region indicating the $d_{\text{Mn-Mn}}$ variation caused by zero-point motion. The red curve denotes the experimentally refined structural path, with the shaded region indicating the uncertainty in $d_{\text{Mn-Mn}}$ from XRD refinement. (b) Full-Brillouin-zone $T_c$ calculations for representative structures selected from panel (a), plotted as a function of $\lambda_\Gamma$. The solid line gives the linear fitting with $R^2$=0.983. (c) $T_c$ and $\lambda_\Gamma$ as functions of $d_{\text{Mn-Mn}}$at V=5.89 Å$^3$/atom, corresponding to 110 GPa in experiments. Blue solid circles show $T_c$ from full Brillouin-zone EPC calculations, and blue open circles show $T_c$ estimated from the linear relation between $T_c$ and $\lambda_\Gamma$ in (b). The red star indicates the experimental $T_c$, with a horizontal bar showing uncertainty in $d_{\text{Mn-Mn}}$ from XRD refinement. (d) Superconducting $T_c$ landscape by applying the $T_c$-$\lambda_\Gamma$ calibration in (b) to the sampled $\lambda_\Gamma$ dataset in (a).

~0.15 Å (Supplemental Material, Text S2), indicating that quantum motion alone can produce variations in the instantaneous Mn-Mn distance comparable to the experimental confidence interval of ~0.10 Å obtained from XRD refinement.

Figure 1(e) also compares the $V$-$d_{\text{Mn-Mn}}$ relation obtained from DFT optimization with that from experimentally refined structures. At the ambient-pressure volume of 7.355 Å$^3$/atom, the equilibrium $d_{\text{Mn-Mn}}$ from DFT differs from the experimental value by only ~0.15 Å, and the confidence intervals overlap. Under compression, however, this offset increases: at ~5.7 Å$^3$/atom, corresponding to ~140 GPa in experiment, the difference reaches ~0.3 Å. Thus, DFT calculations can yield $d_{\text{Mn-Mn}}$ values that differ from experimentally refined structures by 0.1-0.3 Å, with the deviation tending to increase under compression, even after accounting for the uncertainty in $d_{\text{Mn-Mn}}$ from both approaches.

***Electron-phonon coupling landscape.***— To investigate how $V$ and $d_{\text{Mn-Mn}}$ affect EPC, we employ a fast descriptor $\lambda_\Gamma$, which is computed from frozen-phonon calculations and provides an efficient estimate of EPC strength [12]. This descriptor has been shown to correlate well with full Brillouin-zone EPC constants and $T_c$ in boride and hydride systems [12, 13]. We therefore compute $\lambda_\Gamma$ for 561 $MnB_4$ structures densely sampled in the $V$-$d_{\text{Mn-Mn}}$ space, covering both the DFT-relaxed and experimentally refined structural regions. Figure 2(a) shows that $\lambda_\Gamma$ is zero at ambient volume for both the calculated and experimental structures. Upon compression, $\lambda_\Gamma$ remains weak near the DFT-relaxed structural curve but increases markedly toward larger $d_{\text{Mn-Mn}}$ near the experimental $V$-$d_{\text{Mn-Mn}}$ curve. When $V$ is less than 6.20 Å$^3$/atom, the experimental curve and its confidence interval enter the significantly enhanced high-$\lambda_\Gamma$ region.

To calibrate $\lambda_\Gamma$ against the superconducting transition temperature, we selected representative structures from Fig. 2(a) and performed full Brillouin-zone EPC calculations with the Allen-Dynes formula. The resulting $T_c^{\text{AD}}$ show an almost linear correlation with $\lambda_\Gamma$ in Fig. 2(b). We further study the dependence of $\lambda_\Gamma$ and $T_c^{\text{AD}}$ on $d_{\text{Mn-Mn}}$ at a fixed volume of $V$=5.89 Å$^3$/atom, corresponding to 110 GPa in experiment. In Fig. 2(c), both $\lambda_\Gamma$ and the calculated $T_c$ increases rapidly with increasing $d_{\text{Mn-Mn}}$. A change of only ~0.05 Å in $d_{\text{Mn-Mn}}$ increases $T_c$ by about 20 K, indicating that the EPC-induced $T_c$ is extremely responsive to the Mn-Mn dimer distance. We then analyze the electronic and phonon properties associated with this rapid increase. In Supplementary Fig. S3, increasing $d_{\text{Mn-Mn}}$ strongly enhances the number of electronic states at the Fermi level, $N(E_f)$, and softens the screened phonon frequency. Such softening is also a hallmark of a Peierls distortion, which drives the structural transition from the orthorhombic phase to the monoclinic structure in $MnB_4$ [7]. The unscreened phonon frequency changes much less. According to Eq. (2) in the Methods, these trends lead to a significant increase in the zone-center EPC. Consistent behavior is found in the full EPC calculations in Supplementary Fig. S4, where a small increase in $d_{\text{Mn-Mn}}$ softens the relevant phonon mode and increases the phonon linewidth. Thus, subtle changes in the Mn-Mn dimer distance induce coupled electronic and phononic responses that rapidly enhance EPC. Applying the relation between $\lambda_\Gamma$ and $T_c^{\text{AD}}$ in Fig. 2(b) to the full $\lambda_\Gamma$ dataset yields a $T_c$ landscape in the $V$-$d_{\text{Mn-Mn}}$ map in Fig. 2(d). The mechanism identified at fixed volume in Fig. 2(c) extends across the full $V$-$d_{\text{Mn-Mn}}$ map, producing an enhanced-$T_c$ region for structures with smaller volume and larger Mn-Mn distance.

Based on the $T_c$–$V$–$d_{\text{Mn-Mn}}$ in Fig. 2(d), Figure 3(a) quantitatively shows the pressure dependence of the calculated $T_c$ obtained from the $V$–$d_{\text{Mn-Mn}}$ curve solved from experimental XRD. Because $T_c$ is highly responsive to $d_{\text{Mn-Mn}}$, small experimental uncertainties in $d_{\text{Mn-Mn}}$ lead to a broad variation in $T_c$, represented by the green shaded region in Fig. 3(a). The calculated $T_c$ follows a pressure dependence similar to that of the experimental data in Ref. [4]. Specifically, near 5.7 Å$^3$/atom, corresponding to ~140 GPa, the calculated $T_c$ agrees with the experimental value. In the V range from 6.4 to 6.8 Å$^3$/atom, corresponding to 30–60 GPa in experiment, the calculated $T_c$ remains lower than the experimental values by a few kelvins. Overall, the EPC-based $T_c$ shows a pressure evolution consistent with the experiment.

We then analyze the microscopic origin of the EPC in $MnB_4$. Figure 3(b) shows the phonon spectrum for a representative structure with V=5.89 Å$^3$/atom and $d_{\text{Mn-Mn}}$=2.548 Å, corresponding to 110 GPa in experiment. The full Brillouin-zone EPC calculation gives λ=0.63 and $T_c^{\text{AD}}$=14.3 K, in good agreement with the experimental value of 11.4 K at the same volume. The phonon dispersion and Eliashberg spectral function $\alpha^2F(\omega)$ show that the dominant contribution to EPC comes from low-frequency vibrations of Mn atoms. The vibrational eigenvector of the mode with the largest linewidth, shown in Supplementary Fig. S5, is dominated by Mn-Mn dimer stretching. Mapping the frequency of this mode in the $V$–$d_{\text{Mn-Mn}}$ plane further shows systematic softening toward smaller V and larger $V$–$d_{\text{Mn-Mn}}$, the same direction in which $T_c$ are enhanced. Thus, the Mn-Mn dimer distance not only controls the energy landscape but also tunes to the vibrational mode that contributes strongly to the EPC interaction.

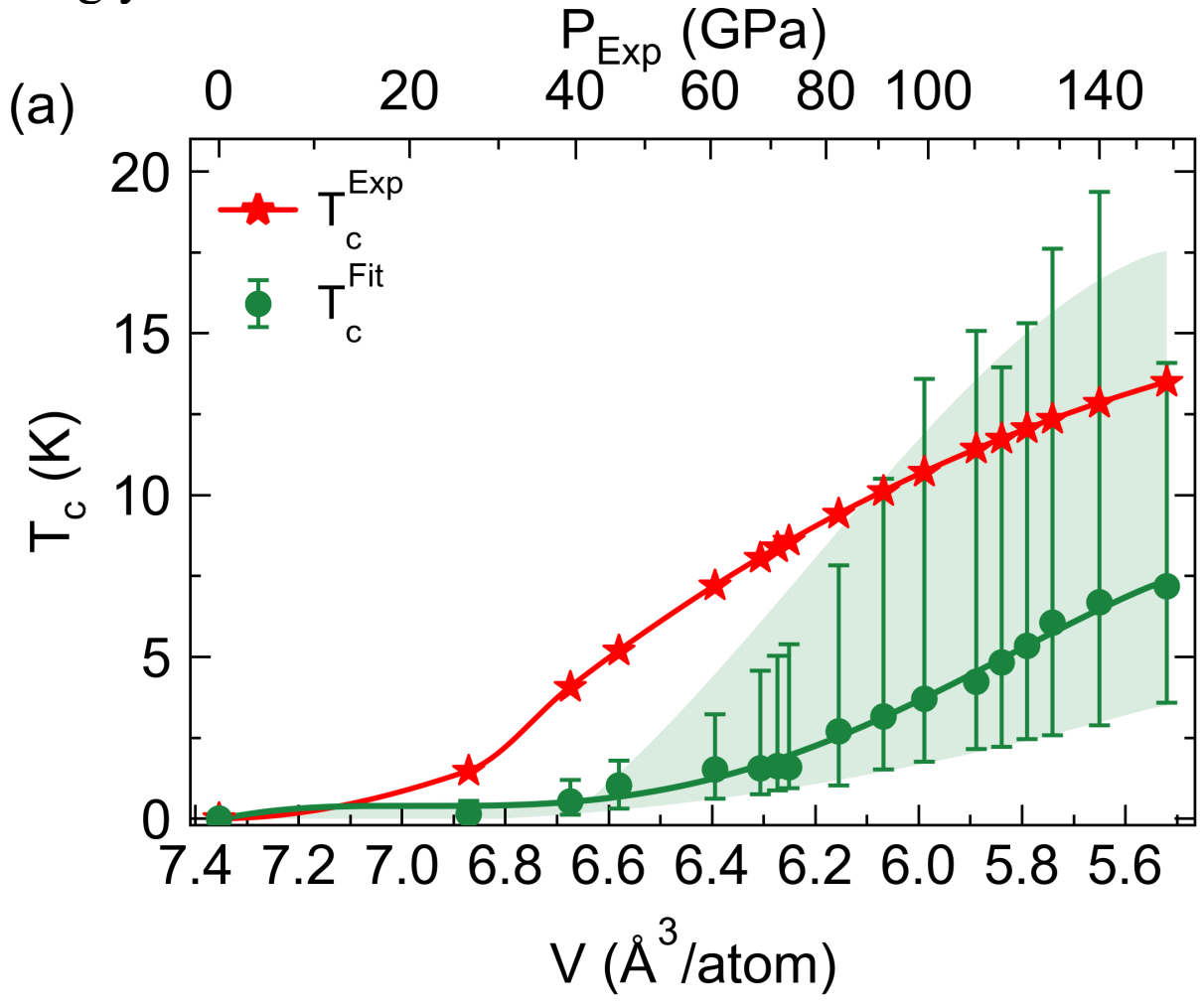

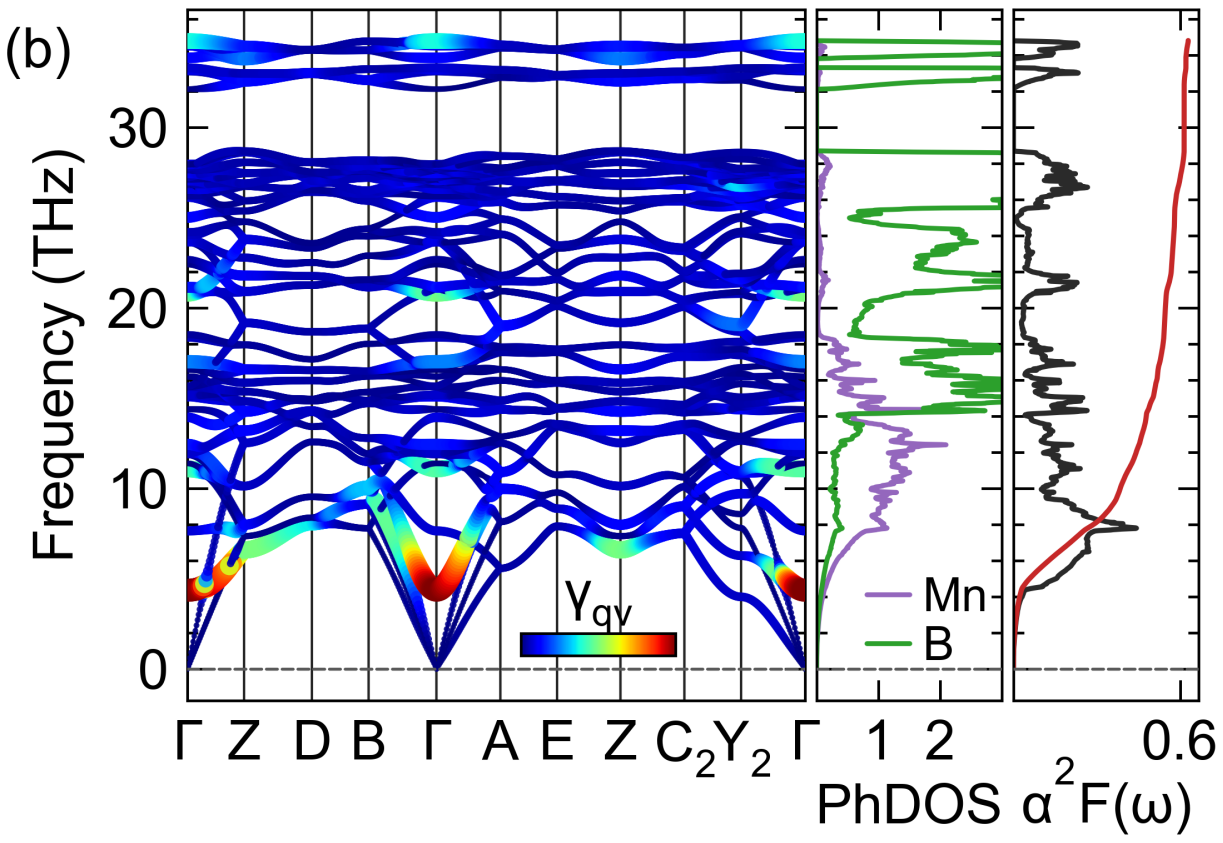

**FIG. 3** Experimental and calculated pressure dependence of $T_c$ in $MnB_4$. (a) Experimental $T_c$ values from Ref. [4] are shown in red. The upper axis shows the pressure corresponding to the experimental equation of state. The green curve gives the calculated $T_c$ obtained from the $T_c$–$V$–$d_{\text{Mn-Mn}}$ map in Fig. 2(d). The green bar and shaded region represent the range of $T_c$ values associated with the experimental uncertainty in $d_{\text{Mn-Mn}}$. (b) Phonon dispersion of $MnB_4$ at $d_{\text{Mn-Mn}}$ = 2.548 Å and $V$ = 5.89 Å$^3$/atom, corresponding to 110 GPa in experiment. The dispersion is shown together with the momentum- and mode-resolved phonon linewidth $\gamma_{\mathbf{q}\nu}$, phonon density of states, Eliashberg spectral function $\alpha^2F(\omega)$, and the cumulative electron-phonon coupling constant $\lambda(\omega)$.

***Magnetic and electronic properties.***—Figure 4(a) shows the magnetic moment $M$ obtained from spin-polarized calculations on the $V$–$d_{\text{Mn-Mn}}$ map. At the ambient-pressure volume (V=7.355 Å$^3$/atom), the magnetic moment is zero for both the DFT-relaxed and experimentally refined structure. This is consistent with previous experiments showing that single-crystal $MnB_4$ is a diamagnetic semiconductor [14], a behavior explained by DFT using the GGA functional [7]. Interestingly, with increasing $d_{\text{Mn-Mn}}$ at this volume, a small finite magnetic moment emerges within the variation range of the experimental $d_{\text{Mn-Mn}}$. This moment is associated with a relatively strongly ferromagnetically coupled Mn-Mn dimer, whereas the magnetic exchange between neighboring dimers remains weak, suggesting possible paramagnetism of these magnetic Mn dimers. This sensitivity of the static magnetic moment to $d_{\text{Mn-Mn}}$ may help explain the higher-temperature magnetism observed in polycrystalline $MnB_4$ samples [5], since thermal fluctuations can naturally increase the variations of $d_{\text{Mn-Mn}}$. In turn, these variations would stabilize spin-fluctuating Mn dimers, leading to observed large paramagnetic moments. Upon compression, both ordered static and fluctuating dynamic moments are strongly suppressed near the experimental $V$–$d_{\text{Mn-Mn}}$ curve. Along the DFT-relaxed $V$–$d_{\text{Mn-Mn}}$ curve, the system remains nonmagnetic over the sampled range.

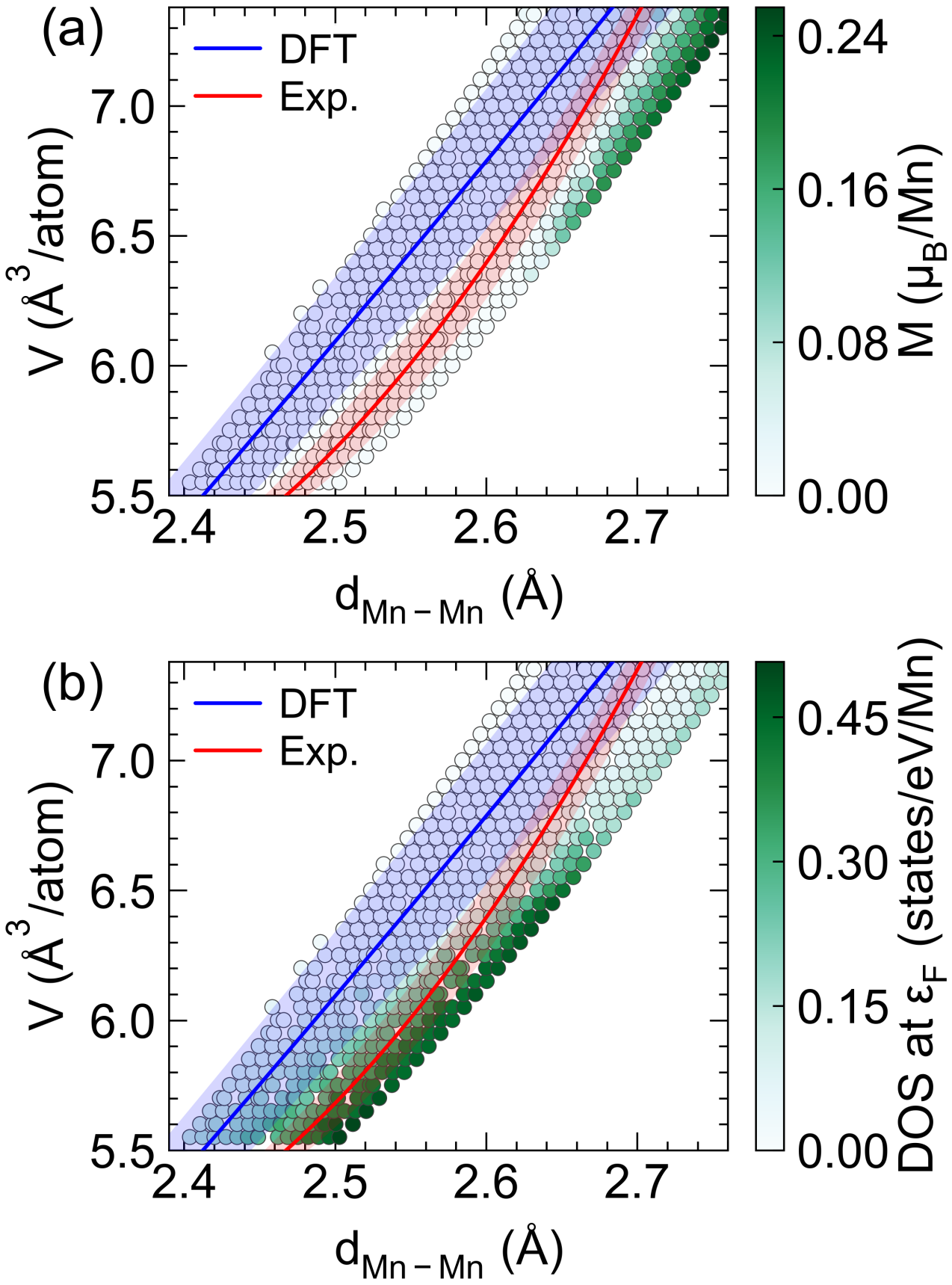

**FIG. 4** Magnetic and electronic properties in the $V$–$d_{\text{Mn-Mn}}$ space. Calculated maps of (a) the Mn magnetic moment and (b) the density of states at the Fermi level $N(\epsilon_F)$ of $MnB_4$ as functions of $V$ and $d_{\text{Mn-Mn}}$. The blue and red curves denote the static 0 K DFT relaxed structure and the experimental structural parameters, respectively.

Figure 4(b) shows the density of states at the Fermi level, $N(\epsilon_F)$. At the ambient-pressure volume, $N(\epsilon_F)$ is zero, consistent with a semiconducting state in the experimental observations [4]. Under compression, $N(\epsilon_F)$ increases strongly along the experimental $V$–$d_{\text{Mn-Mn}}$ curve, indicating an insulator-to-metal transition. At the highest pressures, $N(\epsilon_F)$ is ~0.5 states/eV per Mn, still too small to satisfy the Stoner criterion, which explains the absence of magnetism at high pressure in the calculations. A detailed look at the band structures clarifies how $V$ and $d_{\text{Mn-Mn}}$ modify the electronic band structure in Supplementary Fig. S6. With the same $d_{\text{Mn-Mn}}$, decreasing $V$ pushes dispersive bands toward the Fermi level and closes the semiconducting gap, driving the insulator-to-metal transition under compression. At fixed $V$, increasing $d_{\text{Mn-Mn}}$ reduces the Peierls distortion and produces a similar effect: the bonding-antibonding splitting is reduced, and Mn-$d$ states move closer to the Fermi level, increasing metallicity. As a result, $N(\epsilon_F)$ increases toward smaller $V$ and larger $d_{\text{Mn-Mn}}$ in Fig. 4(b).

***Discussions.***—We have shown that both EPC and magnetism in $MnB_4$ are highly sensitive and responsive to the Mn-Mn dimer distance, which acts as a key structural parameter. Upon compression, EPC remains weak if the calculation follows only the DFT-relaxed structural path. In contrast, using $d_{\text{Mn-Mn}}$ values from the experimentally refined XRD structures and their uncertainty ranges leads to substantially enhanced EPC. The Allen-Dynes formula then yields superconducting transition temperatures comparable to experiment, especially at high pressure (Fig. 3a). In the intermediate-pressure range of 30–60 GPa, however, the calculated $T_c$ remains lower than the experimental values. We find that a further increase of $d_{\text{Mn-Mn}}$ by only ~0.6% would bring the calculated $T_c$ into good agreement with experiment. Given the strong sensitivity of this system to $d_{\text{Mn-Mn}}$, such an additional shift could arise from experimental conditions. Another possible contribution is spin fluctuation, which is not included in the present treatment of superconductivity. This possibility is suggested by the magnetic landscape in Fig. 4(a). Along the experimental $V$–$d_{\text{Mn-Mn}}$ curve, our calculations do not find a robust ordered Mn moment in the superconducting high-pressure region. However, because the magnetic moment is also highly sensitive to $d_{\text{Mn-Mn}}$, a subtle change in the dimer distance can induce weak magnetic moments near the experimentally relevant structural region. This proximity to a magnetic instability suggests that spin fluctuation may coexist with EPC and further affect $T_c$. Together with our findings, these remaining questions could be clarified by future isotope-effect experiments.

The present calculations are based on DFT with the GGA functional, and correlation effects on the electronic structure may be underestimated. To assess this possibility, we use DFT+$U$ to examine whether a Hubbard $U$ modifies the present scenario. Specifically, we recalculate $\lambda_\Gamma$ and the magnetic moment at three fixed volumes for different $d_{\text{Mn-Mn}}$ values, adding $U$ while keeping the crystal structure and other DFT settings unchanged. As shown in Supplementary Fig. S7, at ambient pressure (Fig. S7a), adding $U$ = 1 eV already stabilizes long-range magnetic order for the experimental $d_{\text{Mn-Mn}}$ range. The magnetic moment further increases with larger $U$. Because $MnB_4$ is experimentally nonmagnetic under ambient conditions [14], it indicates that $U$ cannot be large if the nonmagnetic ground state is to be preserved. It also shows that ambient-pressure $MnB_4$ lies close to a magnetic instability, where paramagnetic spin fluctuation may be present. However, its semiconducting behavior at ambient pressure prevents these fluctuations from producing superconductivity. Upon compression, the ordered magnetism is naturally suppressed (Fig. S7b), suggesting that the corresponding paramagnetic spin fluctuation should also weaken. At 147 GPa (Fig. S7c), the nonmagnetic state is already stable for $U$ = 1 eV with the experimental $d_{\text{Mn-Mn}}$. Importantly, both $U$ = 0 and $U$ = 1 eV calculations reproduce the rapid increase of $\lambda_\Gamma$ with increasing $d_{\text{Mn-Mn}}$. Thus, the strong enhancement of EPC with increasing Mn-Mn distance in nonmagnetic $MnB_4$ is a robust effect, even when moderate correlation effects are included.

In summary, we find that the pressure-induced superconductivity of $MnB_4$ can be understood within an EPC framework once its structural response under compression is described in the full $V$–$d_{\text{Mn-Mn}}$ space. Its monoclinic symmetry contains an intrinsic Mn-Mn dimer distortion, so compression can tune both the volume and the degree of dimerization. We therefore treat $V$ and $d_{\text{Mn-Mn}}$ as independent structural variables, explicitly accounting for variations in $d_{\text{Mn-Mn}}$. Mapping $\lambda_\Gamma$ over this enlarged structural space identifies a region of strong EPC, and full Brillouin-zone EPC calculations for representative structures yield $T_c$ values comparable to experimental values. Mode-resolved EPC analysis further shows that phonon modes associated with Mn-Mn dimer stretching provide the main contribution to the enhanced coupling. This enhanced EPC is also supported by an increased density of states at the Fermi level, while remaining insufficient to stabilize a robust magnetic phase. Thus, $d_{\text{Mn-Mn}}$ affects both electronic and phonon properties, leading to the emergence of superconductivity. More broadly, $MnB_4$ represents an HRS, providing a platform to study how subtle changes in an internal degree of freedom affect superconductivity and magnetism. Future isotope-effect experiments should clarify our predictions, and chemical doping in this system may provide another route to study this HRS.

## Methods

***Structure optimization.*****—**First-principles calculations for structural optimization of MnB4 were performed within density functional theory (DFT) using the Vienna Ab initio Simulation Package (VASP) [15, 16] with projector augmented wave (PAW) potentials [17]. Exchange-correlation effects were described using the spin-polarized generalized gradient approximation (GGA) with the Perdew-Burke-Ernzerhof (PBE) functional [18]. A plane-wave cutoff of 520 eV was used, and the total energies were converged to $10^{-5}$ eV. Brillouin zone sampling employed Monkhorst-Pack [19] meshes with a k-point density of $2\pi \times 0.033$ Å$^{-1}$. Both lattice constants and atomic positions were fully relaxed until residual forces on all atoms were less than 0.01 eV/Å.

***Electron-phonon coupling calculation.*****—**The zone-center EPC strength ($\lambda_\Gamma$) is employed as an efficient assessment for the superconductivity which is two orders of magnitude faster than the DFPT calculations [12, 20-22]. The $\lambda_\Gamma$ is computed by

$$\lambda_\Gamma=\sum_\nu \lambda_{\Gamma\nu}, \tag{1}$$

where the summation is over all zone-center vibrational modes. and $\lambda_{\Gamma\nu}$ is defined as

$$\lambda_{\Gamma\nu}=\frac{\tilde{\omega}_{\Gamma\nu}^2-\omega_{\Gamma\nu}^2}{4\omega_{\Gamma\nu}^2}, \tag{2}$$

where $\omega_{\Gamma\nu}$ and $\tilde{\omega}_{\Gamma\nu}$ are screened and unscreened phonon frequencies of mode $\nu$, respectively. Both phonon frequencies were computed using the finite displacement method as implemented in the Phonopy code [12, 23], with a displacement amplitude of 0.03 Å and an energy convergence of $10^{-8}$ eV.

Full Brillouin-zone electron-phonon coupling constants λ and $T_c$ were computed using density functional perturbation theory (DFPT) [24] as implemented in the Quantum ESPRESSO code [25-27] with PBE pseudopotentials from the PSEUDODOJO database [28]. After convergence tests, the plane-wave cut-off and charge density cut-off were set to 100 and 550 Ry, respectively. Self-consistent field (SCF) calculations were performed using a 16×16×16 k-point mesh. DFPT calculations used a 8×8×8 k-point grid and a 2×2×2 q-point grid, with a convergence threshold of $1\times10^{-15}$ Ry. A Gaussian smearing of 0.01 Ry was applied.

The isotropic Eliashberg spectral function was obtained by averaging over the Brillouin zone [29]

$$\alpha^2F(\omega)=\frac{1}{2N(\epsilon_F)}\sum_{\boldsymbol{q}\nu}\frac{\gamma_{\boldsymbol{q}\nu}}{\hbar\omega_{\boldsymbol{q}\nu}}\delta\left(\omega\text{-}\omega_{\boldsymbol{q}\nu}\right), \tag{3}$$

where $N(\epsilon_F)$ is the density of states (DOS) at the Fermi level. $\omega_{\boldsymbol{q}\nu}$ is the phonon frequency of mode $\nu$ at wave vector $\boldsymbol{q}$, and the phonon linewidth $\gamma_{\boldsymbol{q}\nu}$ is defined as $\gamma_{q\nu}=\frac{2\pi\omega_{q\nu}}{\Omega_{BZ}}\sum_{ij}\int d^3k\left|g_{\boldsymbol{k},\boldsymbol{q}\nu}^{ij}\right|^2\delta\left(\epsilon_{\boldsymbol{q},i}\text{-}\epsilon_F\right)\delta\left(\epsilon_{\boldsymbol{k}+\boldsymbol{q},j}\text{-}\epsilon_F\right)$, where $g_{\boldsymbol{k},\boldsymbol{q}\nu}^{ij}$ is the EPC matrix element. $\epsilon_{\boldsymbol{q},i}$ and $\epsilon_{\boldsymbol{k}+\boldsymbol{q},i}$ are eigenvalues of Kohn-Sham orbitals at bands $i$, $j$ and wave vectors $\boldsymbol{q}$, $\boldsymbol{k}$. The integration of the Eliashberg spectral function yields the full Brillouin-zone EPC constant $\lambda$, expressed as

$$\lambda=2\int\frac{\alpha^2F(\omega)}{\omega}d\omega \tag{4}$$

The $T_c$ was estimated using the McMillan equation [30] modified by Allen-Dynes [31, 32] as

$$T_c=\frac{\omega_{log}}{1.2}\exp\left[\frac{-1.04(1+\lambda)}{\lambda(1\text{-}0.62\mu^*)\text{-}\mu^*}\right], \tag{5}$$

where $\mu^*$ is the effective Coulomb pseudopotential, assumed here to be 0.1, and $\omega_{log}$ is the logarithmic average of the phonon frequencies $\omega_{log}=\exp\left[\frac{2}{\lambda}\int\frac{d\omega}{\omega}\alpha^2F(\omega)log\omega\right]$.

**Acknowledgments.—**We thank Prof. Qing Li and Prof. Hai-Hu Wen of Nanjing University for providing the XRD data. The work at Guangdong University of Technology was supported by the National Natural Science Foundation of China (Grant No. 12504069). The work at Xiamen University was supported by the National Natural Science Foundation of China (Grant No. T2422016) and the Natural Science Foundation of Xiamen (Grant No. 3502Z202371007). The work at Jimei University was supported by the National Natural Science Foundation of China (Grant No. 12404077), the Natural Science Foundation of Fujian province (Grant No. 2024J01726), the Natural Science Foundation of Xiamen (Grant No. 3502Z202372015), and the Research Foundation of Jimei University (Grant No. ZQ2023013). V.A. was supported by the U.S. Department of Energy, Office of Basic Energy Sciences, Division of Materials Science and Engineering. Ames National Laboratory is operated for the U.S. Department of Energy by Iowa State University under Contract No. DE-C02-07CH11358.

**References**

[1] L. Bhatt, E. Abarca Morales, A. Y. Jiang, E. K. Ko, Y.-F. Zhao, N. Schnitzer, G. A. Pan, D. Ferenc Segedin, Y. Liu, Y. Yu, C. M. Brooks, A. S. Botana, H. Y. Hwang, J. A. Mundy, D. A. Muller, and B. H. Goodge, Structural modifications in strain-engineered bilayer nickelate thin films, Nature **653**, 76 (2026).

[2] Y. Gao, Robust $A_1$ superconductivity in the kagome lattice, Physical Review B **109**, 214501 (2024).

[3] R. Wang, Y. Sun, V. Antropov, Z. Lin, C.-Z. Wang, and K.-M. Ho, Theoretical prediction of a highly responsive material: Spin fluctuations and superconductivity in FeNiB2 system, Applied Physics Letters **115**, 182601 (2019).

[4] Z.-N. Xiang, Y.-J. Zhang, Q. Lu, Q. Li, Y. Li, T. Huang, Y. Zhu, Y. Ye, J. Sun, and H.-H. Wen, Superconductivity up to 14.2 K in MnB4 Under Pressure, Advanced Materials **37**, 2416882 (2025).

[5] H. Gou, A. A. Tsirlin, E. Bykova, A. M. Abakumov, G. Van Tendeloo, A. Richter, S. V. Ovsyannikov, A. V. Kurnosov, D. M. Trots, Z. Konôpková, H.-P. Liermann, L. Dubrovinsky, and N. Dubrovinskaia, Peierls distortion, magnetism, and high hardness of manganese tetraboride, Physical Review B **89**, 064108 (2014).

[6] A. Knappschneider, C. Litterscheid, N. C. George, J. Brgoch, N. Wagner, J. Beck, J. A. Kurzman, R. Seshadri, and B. Albert, Peierls-Distorted Monoclinic MnB4 with a Mn□Mn Bond, Angewandte Chemie International Edition **53**, 1684 (2014).

[7] Y. Liang, X. Yuan, Y. Gao, W. Zhang, and P. Zhang, Phonon-Assisted Crossover from a Nonmagnetic Peierls Insulator to a Magnetic Stoner Metal, Physical Review Letters **113**, 176401 (2014).

[8] A. C. Larson and R. B. Von Dreele, General structure analysis system (GSAS), Los Alamos National Laboratory Report LAUR 86-748 (2000).

[9] B. Toby, EXPGUI, a graphical user interface for GSAS, Journal of Applied Crystallography **34**, 210 (2001).

[10] P. Haas, F. Tran, and P. Blaha, Calculation of the lattice constant of solids with semilocal functionals, Physical Review B **79**, 085104 (2009).

[11] K. Lejaeghere, G. Bihlmayer, T. Björkman, P. Blaha, S. Blügel, V. Blum, D. Caliste, I. E. Castelli, S. J. Clark, A. Dal Corso, S. de Gironcoli, T. Deutsch, J. K. Dewhurst, I. Di Marco, C. Draxl, M. Dułak, O. Eriksson, J. A. Flores-Livas, K. F. Garrity, L. Genovese, P. Giannozzi, M. Giantomassi, S. Goedecker, X. Gonze, O. Grånäs, E. K. U. Gross, A. Gulans, F. Gygi, D. R. Hamann, P. J. Hasnip, N. A. W. Holzwarth, D. Iuşan, D. B. Jochym, F. Jollet, D. Jones, G. Kresse, K. Koepernik, E. Küçükbenli, Y. O. Kvashnin, I. L. M. Locht, S. Lubeck, M. Marsman, N. Marzari, U. Nitzsche, L. Nordström, T. Ozaki, L. Paulatto, C. J. Pickard, W. Poelmans, M. I. J. Probert, K. Refson, M. Richter, G.-M. Rignanese, S. Saha, M. Scheffler, M. Schlipf, K. Schwarz, S. Sharma, F. Tavazza, P. Thunström, A. Tkatchenko, M. Torrent, D. Vanderbilt, M. J. van Setten, V. Van Speybroeck, J. M. Wills, J. R. Yates, G.-X. Zhang, and S. Cottenier, Reproducibility in density functional theory calculations of solids, Science **351**, aad3000 (2016).

[12] Y. Sun, F. Zhang, C.-Z. Wang, K.-M. Ho, I. I. Mazin, and V. Antropov, Electron-phonon coupling strength from ab initio frozen-phonon approach, Physical Review Materials **6**, 074801 (2022).

[13] S. Chen, Z. Wu, Z. Zhang, S. Wu, K.-M. Ho, V. Antropov, and Y. Sun, High-Throughput Screening for Boride Superconductors, Inorganic chemistry **63**, 8654 (2024).

[14] N. Steinki, J. L. Winter, D. S. Grachtrup, D. Menzel, S. Süllow, A. Knappschneider, and B. Albert, Electronic and magnetic ground state of MnB4, Journal of Alloys and Compounds **695**, 2149 (2017).

[15] G. Kresse and J. Furthmüller, Efficiency of ab-initio total energy calculations for metals and semiconductors using a plane-wave basis set, Computational Materials Science **6**, 15 (1996).

[16] G. Kresse and J. Furthmüller, Efficient iterative schemes for ab initio total-energy calculations using a plane-wave basis set, Physical Review B **54**, 11169 (1996).

[17] P. E. Blöchl, Projector augmented-wave method, Physical Review B **50**, 17953 (1994).

[18] J. P. Perdew, K. Burke, and M. Ernzerhof, Generalized Gradient Approximation Made Simple, Physical Review Letters **77**, 3865 (1996).

[19] H. J. Monkhorst and J. D. Pack, Special points for Brillouin-zone integrations, Physical Review B **13**, 5188 (1976).

[20] R. Wang, Y. Sun, F. Zhang, F. Zheng, Y. Fang, S. Wu, H. Dong, C.-Z. Wang, V. Antropov, and K.-M. Ho, High-Throughput Screening of Strong Electron–Phonon Couplings in Ternary Metal Diborides, Inorganic chemistry **61**, 18154 (2022).

[21] R. Wang, F. Zheng, Z. Zhang, S. Wu, H. Dong, C.-Z.

Wang, V. Antropov, Y. Sun, and K.-M. Ho, Anticorrelation between electron-phonon coupling strength and stability of ternary metal diborides, Physical Review B **111**, 014104 (2025).
[22] F. Zheng, Z. Zhang, Z. Wu, S. Wu, Q. Lin, R. Wang, Y. Fang, C.-Z. Wang, V. Antropov, Y. Sun, and K.-M. Ho, Prediction of ambient pressure superconductivity in cubic ternary hydrides with MH6 octahedra, Materials Today Physics **42**, 101374 (2024).
[23] A. Togo and I. Tanaka, First principles phonon calculations in materials science, Scr. Mater. **108**, 1 (2015).
[24] S. Baroni, S. de Gironcoli, A. Dal Corso, and P. Giannozzi, Phonons and related crystal properties from density-functional perturbation theory, Rev Mod Phys **73**, 515 (2001).
[25] P. Giannozzi, S. Baroni, N. Bonini, M. Calandra, R. Car, C. Cavazzoni, D. Ceresoli, G. L. Chiarotti, M. Cococcioni, and I. Dabo, QUANTUM ESPRESSO: a modular and open-source software project for quantum simulations of materials, Journal of physics: Condensed matter **21**, 395502 (2009).
[26] P. Giannozzi, O. Andreussi, T. Brumme, O. Bunau, M. B. Nardelli, M. Calandra, R. Car, C. Cavazzoni, D. Ceresoli, and M. Cococcioni, Advanced capabilities for materials modelling with Quantum ESPRESSO, Journal of physics: Condensed matter **29**, 465901 (2017).
[27]P. Giannozzi, O. Baseggio, P. Bonfà, D. Brunato, R. Car, I. Carnimeo, C. Cavazzoni, S. d. Gironcoli, P. Delugas, F. F. Ruffino, A. Ferretti, N. Marzari, I. Timrov, A. Urru, and S. Baroni, Quantum ESPRESSO toward the exascale, The Journal of Chemical Physics **152**, 154105 (2020).
[28] M. J. van Setten, M. Giantomassi, E. Bousquet, M. J. Verstraete, D. R. Hamann, X. Gonze, and G. M. Rignanese, The PseudoDojo: Training and grading a 85 element optimized norm-conserving pseudopotential table, Computer Physics Communications **226**, 39 (2018).
[29] G. Eliashberg, Interactions between electrons and lattice vibrations in a superconductor, Sov. Phys. JETP **11**, 696 (1960).
[30] W. McMillan, Transition temperature of strong-coupled superconductors, Physical Review **167**, 331 (1968).
[31] P. B. Allen, Neutron spectroscopy of superconductors, Physical Review B **6**, 2577 (1972).
[32] P. B. Allen and R. Dynes, Transition temperature of strong-coupled superconductors reanalyzed, Physical Review B **12**, 905 (1975).